\documentclass[]{elsart}
\usepackage{graphicx}

\begin{document}
\begin{frontmatter}
\title{Ferromagnetism and triplet superconductivity in the two-dimensional Hubbard model}
\author[1]{Carsten Honerkamp}
\author[2]{and Manfred Salmhofer}
\address[1]{Department of Physics, Massachusetts Institute of Technology, Cambridge MA 02139, USA }  
\address[2]{Theoretische Physik, Universit\"at Leipzig, 
D-04109 Leipzig, Germany}  
\date{\today}
\begin{abstract} 
We review magnetic and superconducting instabilities in the $t$-$t'$ Hubbard model on the two-dimensional square lattice as obtained with functional one-loop renormalization group techniques.  Special emphasis is put on ferromagnetic and triplet superconducting tendencies that could be relevant to the triplet superconductor Sr$_2$RuO$_4$.
\end{abstract}
\end{frontmatter}


The layered perovskite oxide Sr$_2$RuO$_4$ may become 
the first well-established case of triplet superconductivity. 
A series of experiments has provided a quite consistent phenomenological picture of the pairing state\cite{maeno}, and even details like the precise nodal structure of the superconducting gap may soon be known experimentally\cite{maenorio}. 
On the other hand the microscopic understanding of the mechanisms leading to superconductivity is still incomplete. The three-band nature of the low energy electronic spectrum is one source of complications.
However there is a consensus that the superconductivity in Sr$_2$RuO$_4$ arises due to electron-electron interactions.  
Recently substitution of La for Sr has allowed to bring the system close to a ferromagnetic instability\cite{kikugawa}. This is another indication that electronic correlations play a major role in the material.   

Since an understanding of the complete system is a difficult task, 
it may be helpful to reduce the problem somewhat and 
ask if we can find a similar situation of  a vicinity of
triplet superconductivity and ferromagnetism in a one-band Hubbard model 
on the two-dimensional (2D) square lattice. 
The microscopic Hamiltonian we study is 
\begin{eqnarray*} H &=& -t \sum_{\mathrm{n.n.} , \, s } c^\dagger_{i,s} c_{j,s} -t'
\sum_{\mathrm{n.n.n.},\, s } c^\dagger_{i,s} c_{j,s}  +U \sum_i n_{i\uparrow} n_{i 
\downarrow} \, \end{eqnarray*}
with onsite repulsion $U>0$ and hopping amplitudes $t$ and $t'$ between
nearest neighbors (n.n.) and next-nearest neighbors
(n.n.n.) on the 2D square lattice. 
Besides its relevance for the high-$T_c$ cuprates, this model can be considered to describe the so-called $\gamma$-Fermi surface (FS) of Sr$_2$RuO$_4$ which may responsible for both superconductivity\cite{zhitomirsky} and ferromagnetic (FM) tendencies\cite{kikugawa}.

The weak coupling physics of the Hubbard model on the 2D square lattice has been the subject of a large number of studies. Over the last years, functional renormalization group (RG) methods were developed which give a rather detailed and unbiased account of the competition of the various infrared instabilities of the model\cite{zanchi,halboth,honerkamp,tsai}. 
 Recently we have identified a shortcoming of typical one-loop momentum-shell RG methods in the search for ferromagnetism. 
To remedy this we proposed an alternative method, the temperature-flow RG\cite{honerkampfm}. In the following we will briefly describe the results obtained with this method and make some remarks concerning Sr$_2$RuO$_4$. For more details, the reader is referred to Refs. \cite{honerkampfm} and \cite{honrice}.

\begin{figure} 
\begin{center} 
\includegraphics[width=.65\textwidth]{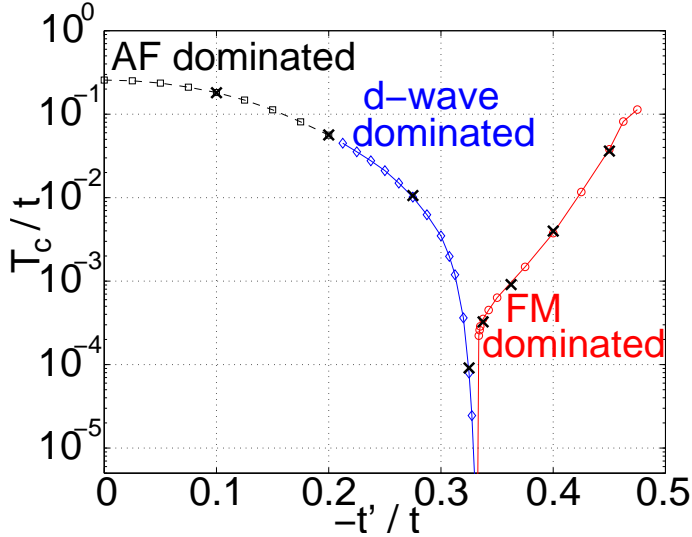}
\end{center}
\caption{
The characteristic temperature for the flow to strong coupling, 
$T_c$, versus next
nearest neighbor hopping amplitude $t'$ from a $T$-flow RG calculation\cite{honerkampfm}. The chemical potential is fixed at
the van Hove value $\mu =4t'$. $T_c$ is defined as temperature where the
couplings reach values larger than $18t$. }
\label{tctp}
\end{figure} 

Here we describe the results for the RG flow with initial interaction $U=3t$ and different values for the next-nearest neighbor hopping $t'$. The chemical potential is adjusted with changing $t'$ such that the FS always contains the van Hove (VH) points at $(\pm \pi,0)$ and $(0,\pm
\pi)$.  The RG flow is summarized in Fig. \ref{tctp}. 
We find three distinct parameter regions. 
The first regime occurs near half-filling for $t'>-0.2t$. 
Here the FS is nested at wave vector $(\pi,\pi)$ and the 
coupling constants flow to strong coupling occurs at relatively high temperatures.  The antiferromagnetic (AF)
susceptibility $\chi_s (\vec{q})$ at $\vec{q}= (\pi,\pi)$  grows most strongly.
When we increase $-t'$, the characteristic 
temperature for the flow to strong coupling $T_c$ decreases. For $t'< -0.2t$, the $d_{x^2-y^2}$-wave pairing susceptibility takes over as the leading susceptibility. This is similar to the findings for $t' \approx 0$ near half filling\cite{zanchi,halboth}. 
Thus there is a wide parameter range for a $d$-wave Cooper instability, i.e. at least at weak coupling the Hubbard model most likely supports superconductivity. By switching off the processes responsible for the growth of the AF susceptibility, we can drastically reduce the superconducting tendencies. This reveals that the $d$-wave superconductivity is induced mainly by AF fluctuations.

When we further increase $|t'|$,
$T_c$ drops rapidly for $t' \le -0.33t$, while for $t' \ge -0.33t$ it rises again. 
Now the flow is dominated by processes with small momentum transfer 
and the FM susceptibility $\chi_s (\vec{q} = 0)$ is by far the most divergent
susceptibility at low temperatures. 
The suppression of $T_c$ around $t'\approx -t/3$ suggests a quantum critical point between $d$-wave singlet superconductivity and ferromagnetism. 
T-matrix approximation (TMA) applied by Hlubina et al. \cite{hlubina}
indicated a FM state beyond a critical $t' \approx
-0.43t$. Recently Irkhin et al.\cite{irkhin} and Hankevych et al.\cite{hankevych} obtained results quite similar to ours.
\begin{figure} 
\begin{center} 
\includegraphics[width=.65\textwidth]{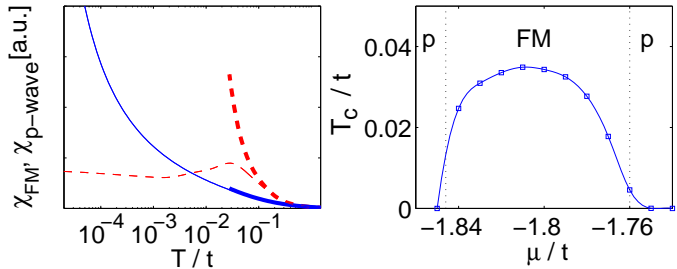}
\end{center}
\caption{Left:
Flow of ferromagnetic (solid lines) and $p_x$-wave (dashed lines) susceptibilities
 for $t'=-0.45t$ at $\mu=-1.78t$ (thick lines) and $\mu=-1.74t$ (thin
lines). Right: Characteristic
temperature $T_c$  versus
chemical potential $\mu$. At the van Hove filling $\mu=-1.8t$.  
}
\label{pwtp45}
\end{figure} 

\begin{figure} 
\begin{center} 
\includegraphics[width=.65\textwidth]{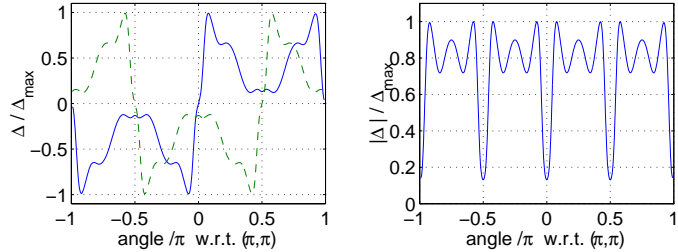}
\end{center}
\caption{Left: Angular variation of real (solid line) and imaginary
(dashed line) part of the gap function obtained with the pair scattering from the RG. Right: Angular variation of the gap magnitude. For these data,
$\mu=-1.71t $ and $t'=-0.45t$.}
\label{gapeq}
\end{figure}

Next we investigate the flow on the FM side for $t'<-t/3$ away from the VH filling. As shown in the right plot of Fig. \ref{pwtp45} for $t'=-0.45t$, the critical  temperature drops by several orders of magnitude when we increase the particle density from the VH value $\langle n \rangle \approx 0.47$ per site at $\mu=-1.8t$ to $\langle n \rangle \approx 0.58$ at $\mu=-1.7t$. 
Furthermore, upon moving away from the VH filling, the growth of the
FM susceptibility gets cut off and the $p$-wave triplet superconducting
susceptibilities with symmetry $p_x\propto \cos \theta$ or $p_y\propto \sin \theta$ diverge at low
temperature. 
Again, very similar to the interplay between AF and $d_{x^2-y^2}$
superconducting fluctuations for small $|t'|$, 
on the one-loop level there is a smooth evolution of the flow to strong coupling from the FM dominated to the $p$-wave dominated instability. However our analysis cannot give a detailed picture of the transition.
It is most plausible to assume that the superconducting ground state will be given by a time-reversal symmetry-breaking superposition of the two components with symmetries $p_x$ and $p_y$, as this maximizes the condensation energy.
Due to symmetry reasons the gap magnitude is strongly suppressed near the VH singularities at $(\pi,0)$ and $(0,\pi)$ (see Fig. \ref{gapeq}). Thus the $p$-wave state cannot benefit from the large density of states in these Brillouin zone regions and occurs at a much lower temperature scale than the $d$-wave state for small $t'$.

In conclusion, regarding the physics of Sr$_2$RuO$_4$ the weak coupling RG analysis of the one-band $t$-$t'$ Hubbard model on the two-dimensional square lattice reveals the following. There is strong evidence for a triplet superconducting instability for hopping amplitudes $-t/2 < t'<-t/3$ in the vicinity of the van Hove filling. Although the in-plane gap function may be gapless, there are symmetry constraints that suppress the gap magnitude near the VH points $(\pi,0)$ and $(0,\pi)$. Thus the energy scale for triplet superconductivity is much smaller than the one for singlet pairing.
Furthermore, for a FS very close to the VH points, the leading instability is in the ferromagnetic channel. This theoretical picture is in qualitative agreement with recent experiments on La-doped Sr$_2$RuO$_4$ which suggest an incipient FM instability\cite{kikugawa}.

We thank T.M. Rice and M. Sigrist for many useful discussions. The German Research Foundation (DFG) is acknowledged for financial support.

 \end{document}